\documentclass[10pt,twocolumn,letterpaper]{article}

\usepackage{cvpr}
\usepackage{times}
\usepackage{epsfig}
\usepackage{graphicx}
\usepackage{amsmath}
\usepackage{amssymb}

\usepackage{lipsum}
\usepackage{mathtools}
\usepackage{cuted}
\usepackage{float}
\usepackage{color}
\usepackage{subfigure}
\usepackage{lscape}
\usepackage{booktabs}
\usepackage{amsfonts}
\usepackage{amsmath,bm}
\usepackage[intoc]{nomencl}
\usepackage{url}
\usepackage[vlined,linesnumbered,boxruled]{algorithm2e}
\usepackage{relsize}
\usepackage{commath}
\usepackage{float}
\usepackage{textcomp}
\usepackage{array}
\usepackage[nottoc,notlot,notlof]{tocbibind}
\usepackage{setspace}
\usepackage{placeins}

\usepackage{hyperref}
\hypersetup{breaklinks=true,bookmarks=false}

\cvprfinalcopy 

\def\cvprPaperID{****} 

\begin{document}

\title{Deep Reinforcement Learning for Field Development Optimization}

\author{Yusuf Nasir \\ 
Energy Resources Engineering \\ 
Stanford University \\ 
{\tt\small nyusuf@stanford.edu}
}

\maketitle

\begin{abstract}
    The field development optimization (FDO) problem represents a challenging mixed-integer nonlinear programming (MINLP) problem in which we seek to obtain the number of wells, their type, location, and drilling sequence that maximizes an economic metric. Evolutionary optimization algorithms have been effectively applied to solve the FDO problem, however, these methods provide only a deterministic (single) solution which are generally not robust towards small changes in the problem setup. In this work, the goal is to apply convolutional neural network-based (CNN) deep reinforcement learning (DRL) algorithms to the field development optimization problem in order to obtain a policy that maps from different states or representation of the underlying geological model to optimal decisions. The proximal policy optimization (PPO) algorithm is considered with two CNN architectures of varying number of layers and composition. Both networks obtained policies that provide satisfactory results when compared to a hybrid particle swarm optimization - mesh adaptive direct search (PSO-MADS) algorithm that has been shown to be effective at solving the FDO problem.
\end{abstract}

\section{Introduction}

The oil and gas industry is faced with the dual energy challenge of extracting oil and gas in a cost effective manner and also seeking ways to reduce the carbon footprint. This presents challenging subsurface optimization problems that appear in different forms. On one hand, we have the field development optimization (FDO) problem where decisions on the number of wells, their type (extraction/injection), location, drilling sequence, and well operational settings need to be made to optimally extract hydrocarbon in order to maximize an economic metric. On the other hand, there is the optimization problem related to carbon-capture and sequestration (CCS) where the goal is to optimally find 'sites' where carbon dioxide ($CO_{2}$) can be optimally injected and stored to prevent leakage from the subsurface in the future.

The goal of this work is to investigate the performance of convolutional neural network-based (CNN) deep reinforcement learning (DRL) algorithm to the field development optimization problem (with fixed well operational settings). In the field development problem, the production life-cycle is divided into a number of discrete drilling stages. Our goal in each stage is to decide if to drill a well or not. If the optimal decision is to drill a well,  the decision on the well type and well location will also need to be made.  This is a very challenging mixed-integer nonlinear programming (MINLP) problem that requires a significant number of expensive flow simulations. Although the field development optimization problem is considered in this work, the procedures to be developed could be applicable to the carbon capture and sequestration (CCS) problem. 

Several approaches have been proposed to solve different variants of the field development problem. Evolutionary optimization algorithms, such as particle swarm optimization \cite{onwunalu2010application, isebor2014a}, genetic algorithm \cite{montes2001use} etc., have been applied to the field development optimization problem. As the number of drilling stages increases, the performance of these algorithms deteriorate, due to increase in the number of decision variables. Also they provide a single solution for the specific problem definition and hence cannot be generalized. Thus, for a small change in the optimization setup, a new optimization problem need to be solved (although previous solutions can be used to improve efficiency). However, if an optimal policy is obtained using DRL, it might be robust to small changes in the optimization setup. 

For this reason, the proximal policy optimization (PPO) \cite{schulman2017proximal} algorithm that have been shown to perform satisfactorily for problems with large action/decision spaces and easily parallelizable is proposed to solve the field development optimization problem in this work. The performance of PPO is investigated using two different CNN architectures of varying number of layers and composition. An efficient hybrid derivative-free optimization strategy that combines particle swarm optimization (PSO) and mesh adaptive direct search (MADS), for global exploration and local exploitation of the search space will be used to benchmark the performance of the CNN-based DRL algorithm. Specifically, the computational cost and objective function value (net present value, NPV in this work) will be compared for the different algorithms.

\section{Related Work}

There have been a number of studies that have applied different optimization algorithms to some aspects of the field development optimization problem. Gradient-based optimization approaches \cite{brouwer2002dynamic, sarma2006efficient} have been used to efficiently optimize the well control/operational settings (parameterized by continuous variables) with fixed well locations. Other work have considered the separate well placement optimization problem \cite{onwunalu2010application, bouzarkouna2012well} and the joint optimization of both the well placement and control problem \cite{bellout2012joint, isebor2014a, nasir2019hybrid}. Due to the presence of multiple local optima in the search space of the well placement problem, derivative free (DF) approaches have been shown to provide better performance. DF approaches also provide satisfactory results when the number of wells, their type and drilling sequence are considered in the optimization \cite{isebor2014a}. As noted earlier, in all of these studies, the goal is to obtain a single solution that maximizes an objective function, typically the net present value (NPV).

Reinforcement learning algorithm have been applied to solve the well control optimization problem \cite{hourfar2019reinforcement}. Deep reinforcement learning has also been applied to the well control optimization problem \cite{ma2019waterflooding, miftakhov2020deep}.  In \cite{ma2019waterflooding}, they evaluated the performance of deep Q-network (DQN), double DQN (DDQN), dueling DDQN, and deep deterministic policy gradient (DDPG) on the well control optimization problem with fully connected neural networks (FCNN). The state space which are two-dimensional maps are flattened and used as input and thus not retaining the spatial information inherent in the data. An FCNN-based PPO was also used in \cite{miftakhov2020deep}.

The focus in this work is on the more challenging field development optimization problem in which the decision to be made include the number of wells, their locations, types and drilling sequence. A CNN-based PPO is considered in order to retain the spatial information present in the observation space (dynamic state and static maps). 

\section{Optimization Problem and Data}

 Our goal in this work is to use DRL to find a policy ($\pi$) that maps from states (representation of the environment/reservoir model) to optimal actions/decisions for the field development problem. In general, the DRL problem can be represented as follows:

\begin{gather}
\begin{array}{rrclcl}
\displaystyle \max_{\theta} & {J (\pi(a \ | \ s, \ \theta))}
\end{array}
\label{eq:drl}
\end{gather}

\noindent 
where $J$ is the cumulative reward/objective function to be optimized, $\theta$ is the weights of the neural network that defines the policy. The variables $s$ and $a$ are the states and actions, respectively. The policy $\pi(a \ | \ s, \ \theta)$ defines the action conditioned on the neural network weight and environment (reservoir model) state.

In this work, the cumulative reward $J$ is defined as the net present value (NPV). Following \cite{Shirangi2015Closed-LoopValidation}, we compute NPV as follows:

\begin{multline}
	\textnormal{NPV($\pi$)}=\sum\limits_{k = 1}^{N_t}  \left[ \sum\limits_{i = 1}^{N_p} \left(p_{o} ~q^{i}_{o,k}-c_{pw}~q^{i}_{pw,k}\right) - \sum\limits_{i = 1}^{N_i} c_{iw}~q^{i}_{iw,k} \right]\times\\
	\frac{\Delta t_k}{( 1+b )^{t_k/365}} - \sum\limits_{i = 1}^{N_w}  \frac{|w_i|~c_w}{(1+b)^{ t_i/365}}.
    \label{gen_field_dev_npv_eqn}
\end{multline}
\noindent 
Here $N_i$ and $N_p$ are the number of injection and production wells, respectively, $N_t$ is the number of time steps in the flow simulation, $t_k$ and $\Delta t_k$ are the time and time step size at time step $k$, $t_i$ is the time at which well $i$ is drilled, and $p_{o}$, $c_{pw}$, and $c_{iw}$ represent the oil price and the cost of produced and injected water. The variables $p_{o}$, $c_{pw}$, and $c_{iw}$ are set to \$55/STB, \$6/STB, and \$2/STB, respectively. The variables $c_w$ and $b$ represent the well drilling cost and annual discount rate which are set to \$25 million and 8\% respectively. The rates of oil/water production and water injection, for well $i$ at time step $k$ are, respectively, $q^{i}_{o,k}$, $q^{i}_{pw,k}$, and $q^{i}_{iw,k}$. Similar equation can be written for a specific drilling stage by considering only the production, injection and drilling information at that stage.

In this work, the Stanford's Automatic Differentiation-based General Purpose Simulator (AD-GPRS) \cite{Zhu_ADGPRS} is used for the flow simulation to obtain quantities from which the objective function is computed and also the dynamic state maps (described later) that are components of the state space. A restart strategy is used to improve the efficiency of the overall flow simulation. This entails saving information of previous flow simulations (due to actions taken in previous drilling stages) that allows for the simulation to be restarted at the beginning of the next drilling stage without the need to run the flow simulation from scratch. In this work, five drilling stages are considered each of length 150 days.

\begin{figure*}[!htb]
    \centering
    \subfigure[Permeability map (in log scale)] {\label{fig:ds2}\includegraphics[width=0.35\textwidth]{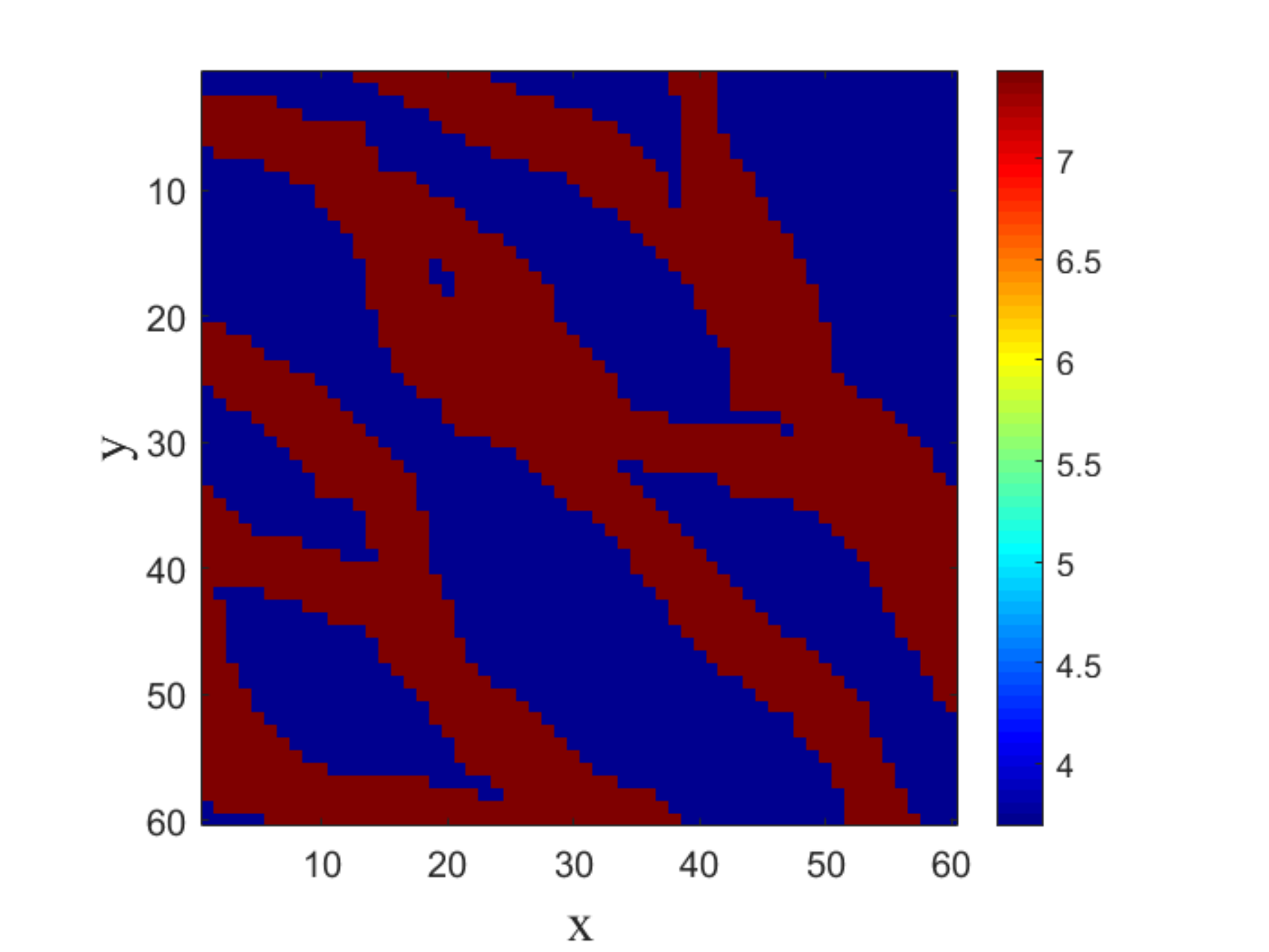}}
    \subfigure[Pressure map] {\label{fig:ds3}\includegraphics[width=0.35\textwidth]{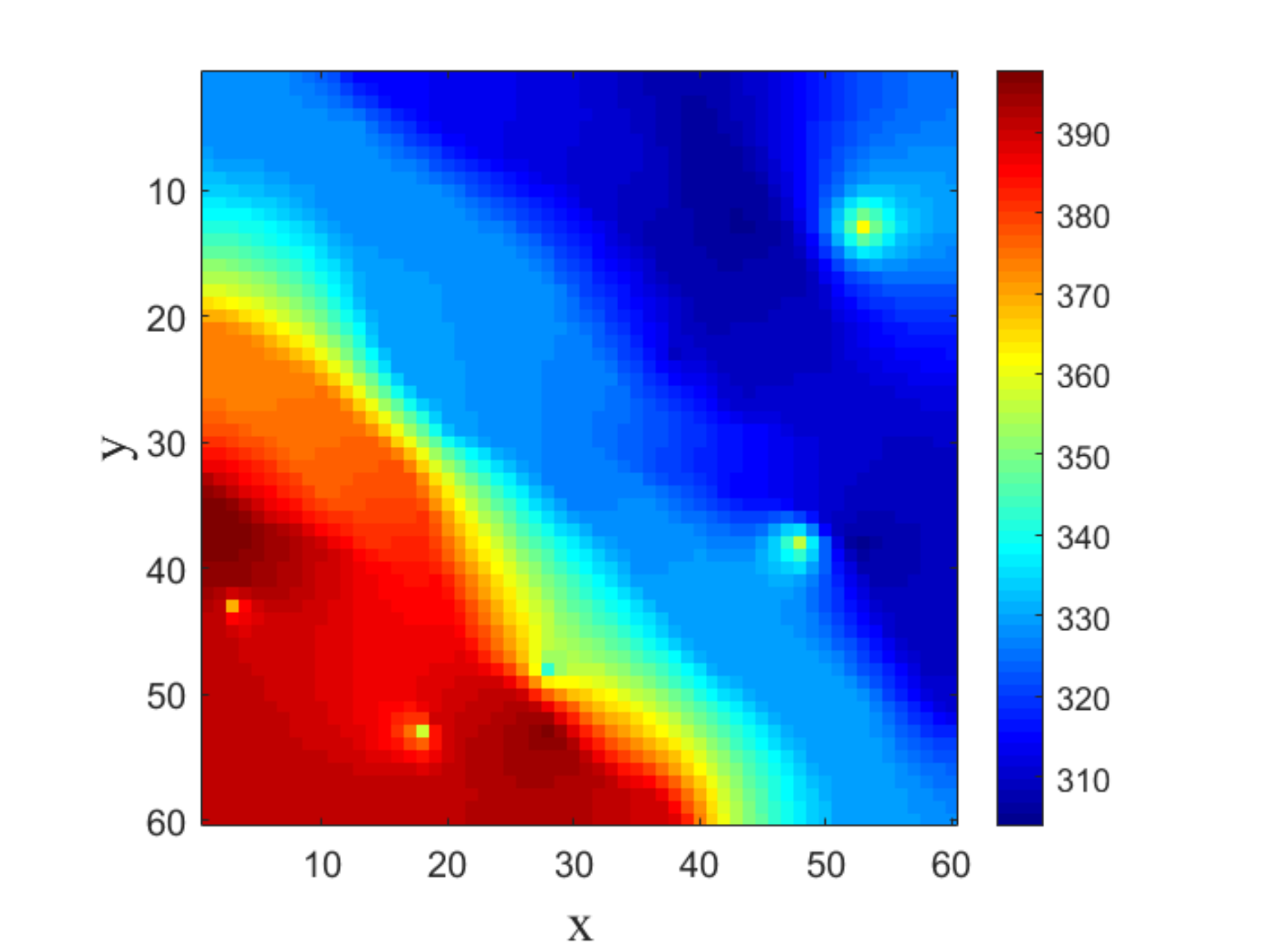}}
	\subfigure[Saturation map] {\label{fig:ds3}\includegraphics[width=0.35\textwidth]{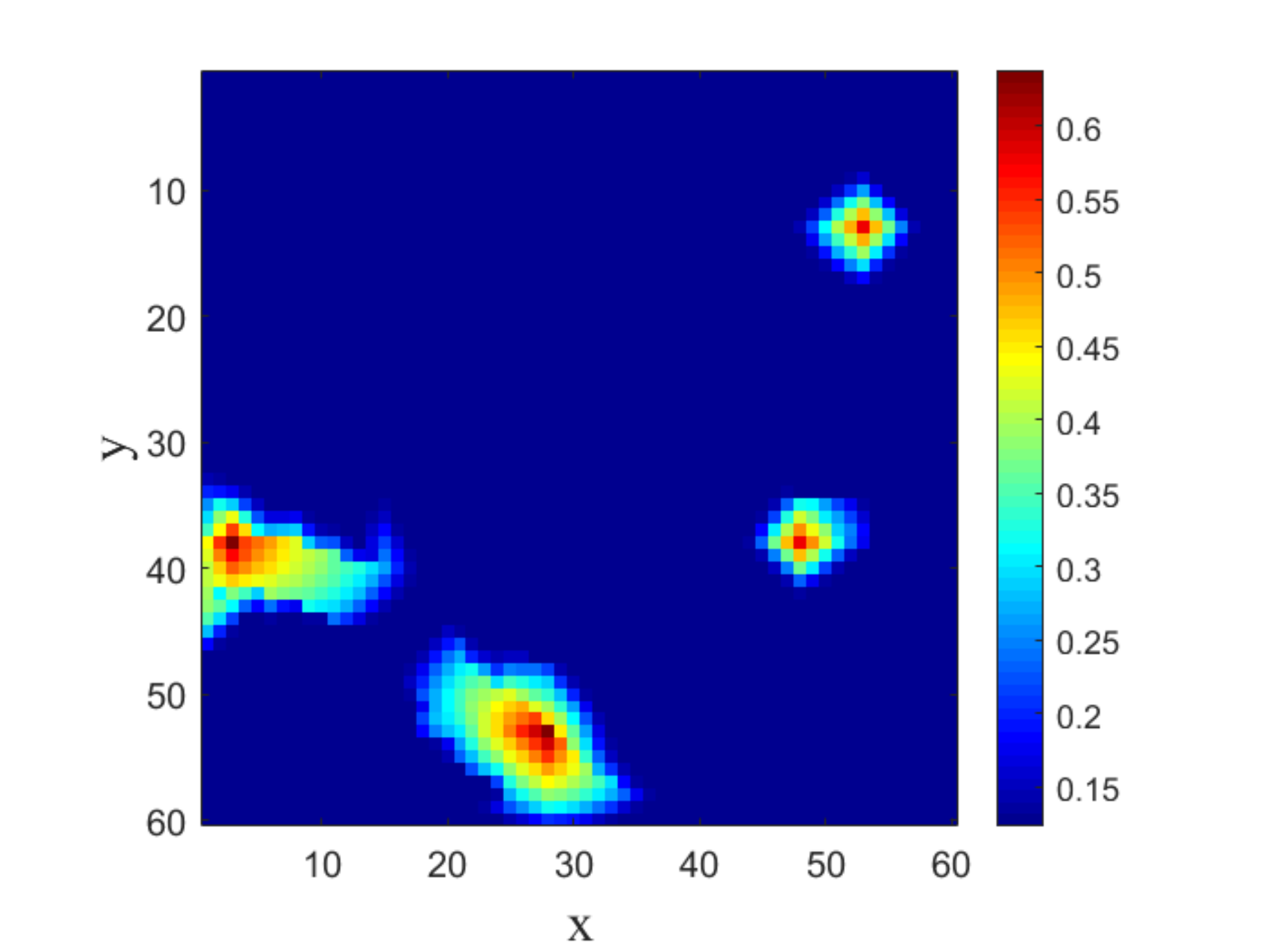}}
	\subfigure[Well location map] {\label{fig:ds3}\includegraphics[width=0.35\textwidth]{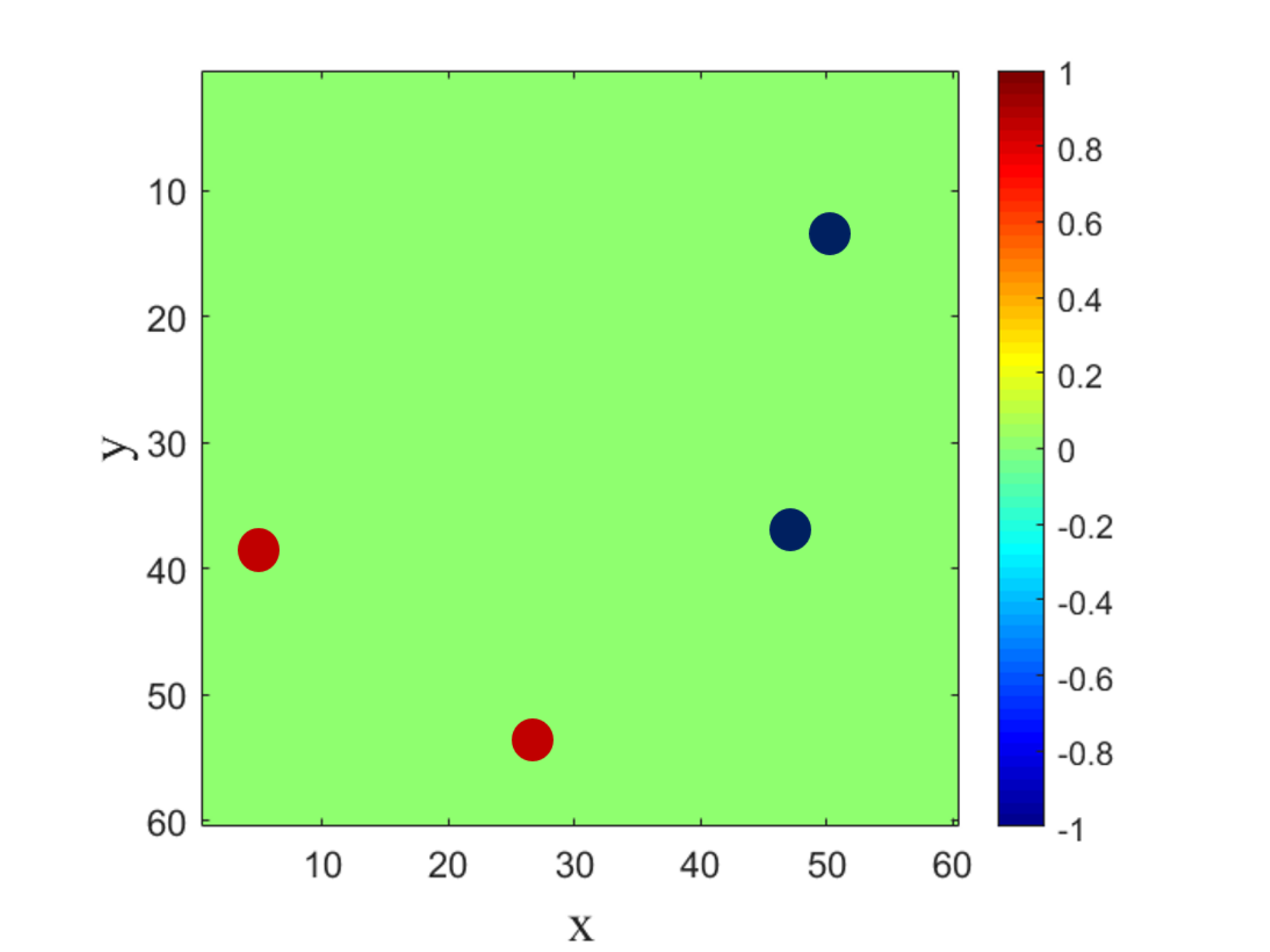}}
    \caption{The four 2D maps that make op the map component of the state space. The red circles in the well location map denotes injectors (1) and blue denotes producers (-1).}
	\label{fig:obs}
\end{figure*}

\section{Methodology}

In this section, the action and observation space are described. Finally, the CNN and PPO DRL algorithm are also discussed.

\subsection{Action and state space}

The key components of deep reinforcement learning includes the action, observation (or state) spaces, and reward. The reinforcement learning agent performs certain actions in the environment. This results in a new state for the environment and a reward signal that indicates the quality of the action. The action and observation spaces for the FDO problem are now described. In the action space, the well type and drill/do not drill decision at drilling stage $k$ is represented by a ternary variable $w_{k} \in \{-1,0,1 \}$ where -1 represents the decision to drill a producer, 1 the decision to drill an injector, and 0 the decision to not drill the well. This is however implemented with a shift of one in the OpenAi gym environment \cite{brockman2016openai}  because it does not take on negative values. The second action (active only when the agent decides to drill a well) which represents where the well should be drilled is represented by a discrete variable $\textbf{u} \in \mathbb{Z}^{N_xN_y}$, where $N_x$ and $N_y$ represents the number of grid blocks in the areal $x$ and $y$ directions. The reservoir model in this work has $N_x = N_y = 60$. Fig. \ref{fig:obs} shows a typical representation of the state space maps at a particular stage of the development.

At each drilling stage, the state/observation space is represented by both two-dimensional maps and a vector. The 2D maps in the state space include the permeability, pressure, saturation and location maps. The permeability map is a static (in this work) measure that indicates the conductivity of fluid in different section of the subsurface model. The dynamic pressure and saturation maps are generated by the reservoir simulator. They depend on the action taken by the agent. For example, the regions in which an injection well is drilled (for pressure support) will have a relatively higher pressure than regions where a production well is drilled. Accordingly, since water is injected, the injection regions should have high water saturation. The well location map indicates the location of wells that have already been drilled and their type. 

\begin{figure*}[!htb]
    \centering
    \subfigure[Small network] {\label{fig:small_net}\includegraphics[width=0.35\textwidth]{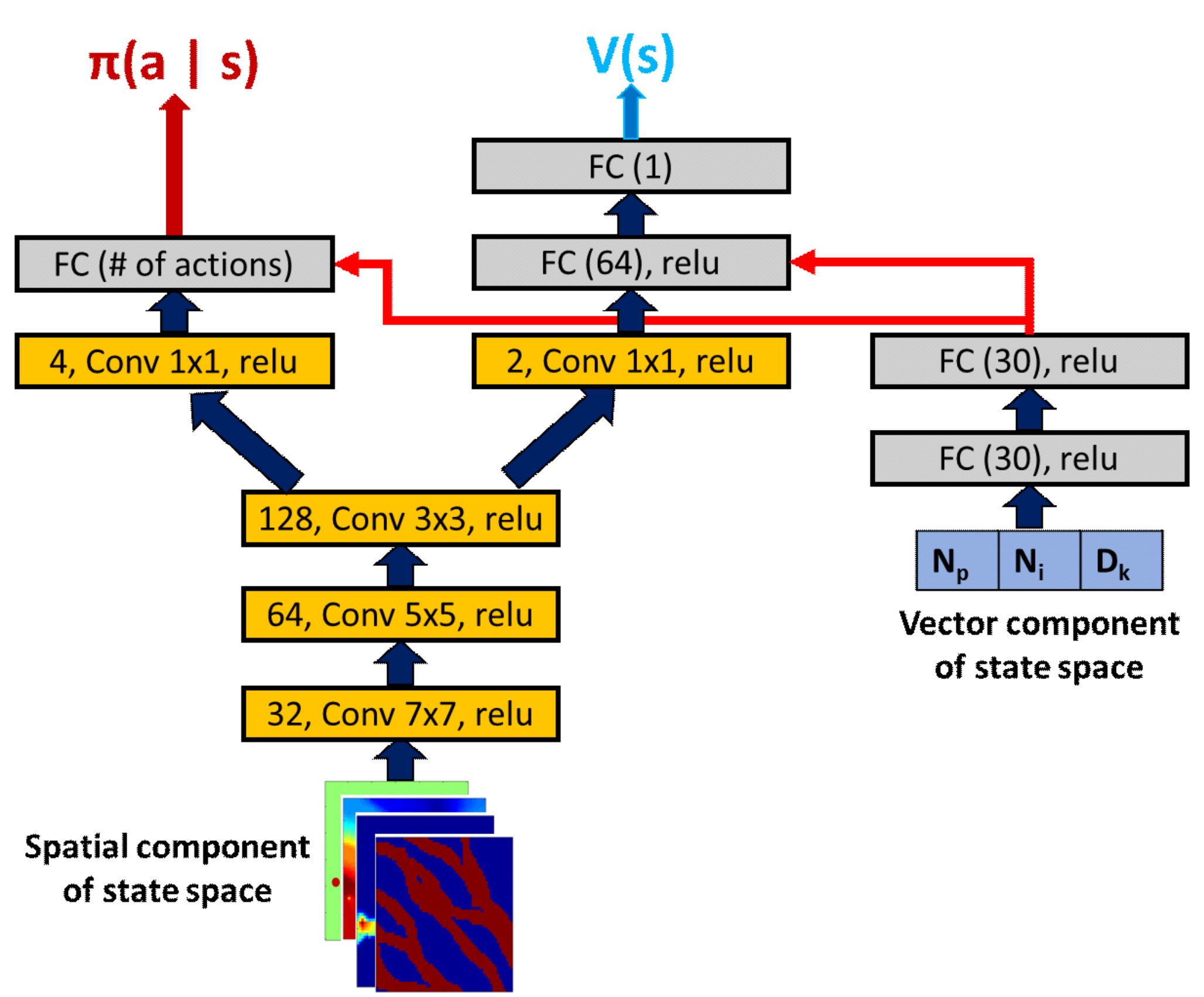}}
    \hspace{15mm}
    \subfigure[Large network] {\label{fig:big_net}\includegraphics[width=0.35\textwidth]{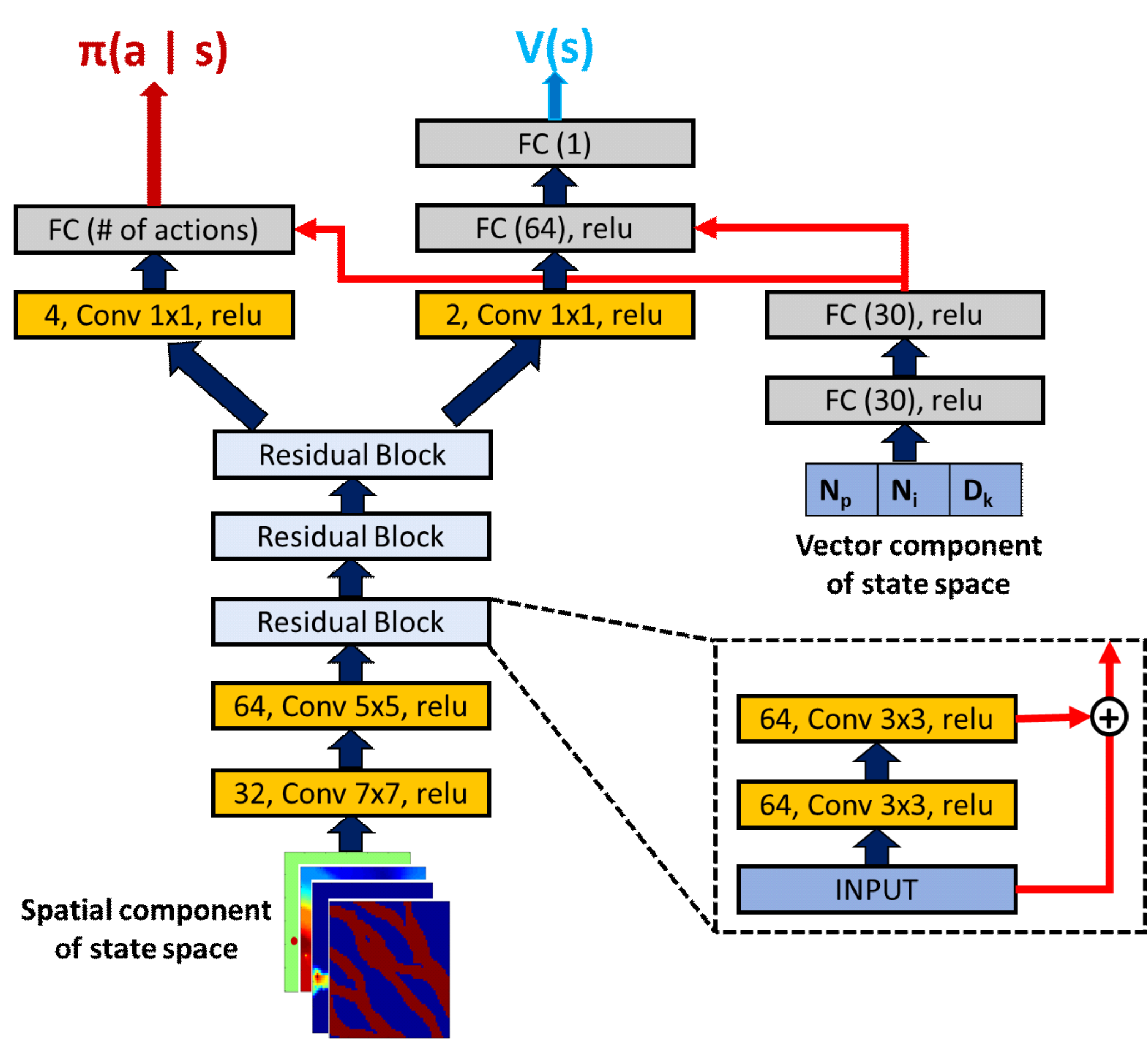}}
	
    \caption{Small and large CNN architectures. The first number in the convolutional layers represent the number of filters and the number in parenthesis of the fully connected layers denote the number of neurons. All convolution are of stride one with no padding (except in residual blocks with same padding).}
	\label{fig:cnn_arc}
\end{figure*}

The permeability and pressure map are normalized to values between 0 and 1. The saturation which is a percentage is given between 0 and 1 and hence do not need to be normalized. The well location map however is more like a mask with zeros and have value of -1 where a producer is drilled and 1 where an injector is drilled. Although the effect of the use of numbers on opposite side of the number line is not isolated and studied, the goal is to allow the agent to easily capture the opposite effects of producing and extracting wells (pressure decrease and increase). The spatial component of the state space $\textbf{s}_{m} \in \mathbb{R}^{N_y \times N_x \times c}$, where $c$ defines the number of 2D maps ($c = 4$ in this work). 

The vector component of the state space at each drilling stage $\textbf{s}_{v} \in \mathbb{R}^{3}$ contains the normalized (between zero and one) drilling stage number, number of producers and injectors that have been drilled at that stage. The number of producer and injectors which are key quantities in FDO are explicitly defined in the vector component, even though they are implicitly defined in the well location map, in order to accelerate learning. Following \cite{pardo2017time} that showed the importance of incorporating time in reinforcement learning problems, the drilling stage number (which also defines the remaining time in the finite horizon field development problem) was included in the vector component of the observation space in order to satisfy the Markov property.

\subsection{DRL algorithm and CNN architecture}

As noted earlier, in this work the proximal policy optimization (PPO) \cite{schulman2017proximal} deep reinforcement learning algorithm is considered. PPO is an on-policy DRL algorithm that can be used with either discrete or continuous action spaces. The PPO with clipped surrogate objective (PPO-clip) is used in this work. The policy update for PPO-clip using multiple steps of stochastic gradient descent is given by:

\begin{gather}
\begin{array}{rrclcl}
\displaystyle \theta_{k+1} =  arg \max_{\theta} & \underset{s, a\sim \pi _{\theta_{k}}}{\mathbb{E}} \left [{L (s, a , \theta_{k}, \theta)}\right]
\end{array}
\label{eq:field_dev_opt_eqn}
\end{gather}

\noindent
where $L$ is given by

\begin{multline}
    \displaystyle L (s, a , \theta_{k}, \theta) =  \min \left( \frac{\pi_{\theta}(a | s)}{\pi_{\theta_{k}}(a | s)} A^{\pi_{\theta_{k}}}(s,a), \right. \\  \left. \left(\frac{\pi_{\theta}(a | s)}{\pi_{\theta_{k}}(a | s)}, \ 1 - \epsilon, \ 1 + \epsilon \right) A^{\pi_{\theta_{k}}}(s,a)) \right)
    \label{eq:clip2}
\end{multline}

\noindent
Here, $\epsilon$ is a hyperparameter that determines the trust region (around the old policy) during the policy update and the new policy is not allowed to go outside this region. The advantage function, $A$, defines the 'quality' of an action without considering the 'quality' of its state. In addition to the policy output the defines the probability distribution of the action space at a specific state, the agent will need to compute the value/quality of the state $V(s)$ in order to calculate the advantage function. RLLib's \cite{liang2017rllib} implementation of PPO-clip is used in this project.

Two CNN architectures of varying number of layers and composition, shown in Fig. \ref{fig:cnn_arc}, are considered. In the small network shown in Fig. \ref{fig:small_net}, the state maps are convolved through three convolutional layers. The output of the third layer is shared by the policy and value arms of the network. The vector components of the state space which include the drilling stage index ($D_k$), number of producers($N_p$) and injectors($N_i$), are processed by two fully connected layers. The output of the second fully connected layer is concatenated (individually) with both the output of the first convolution in the value and policy arms of the network. The individual concatenated tensors are further processed by fully connected layers to produce the probability distribution of the action space condition on the state space ($\pi(a | s)$ for the policy network) and a single digit that indicated the value of the state ($V(s)$ for the value network). The large network shown in Fig. \ref{fig:big_net} is similar to the small network except three residual blocks are introduced to replace the last convolution in the shared layers by the policy and value arms. 

A sample size of 320 episodes (full flow simulation) is specified to collect experiences by rolling out the policy until the terminal state is reached using 40 CPU processors. This means the effective computational cost (neglecting IO processing for reading and saving restart files) to collect experiences at each iteration is the cost of eight flow simulations. Adam optimizer is used with an initial learning rate of 0.001 and a batch size of 160 episodes is used for training. A single Nvidia V100 GPU is used to train the CNN network. The RLlib package \cite{liang2017rllib} handles the communication between the GPU and CPUs and the parallel execution of the flow simulation with the CPUs using a custom field development gym environment developed in this work.

\begin{figure}
\centering
\includegraphics[width=0.5\textwidth]{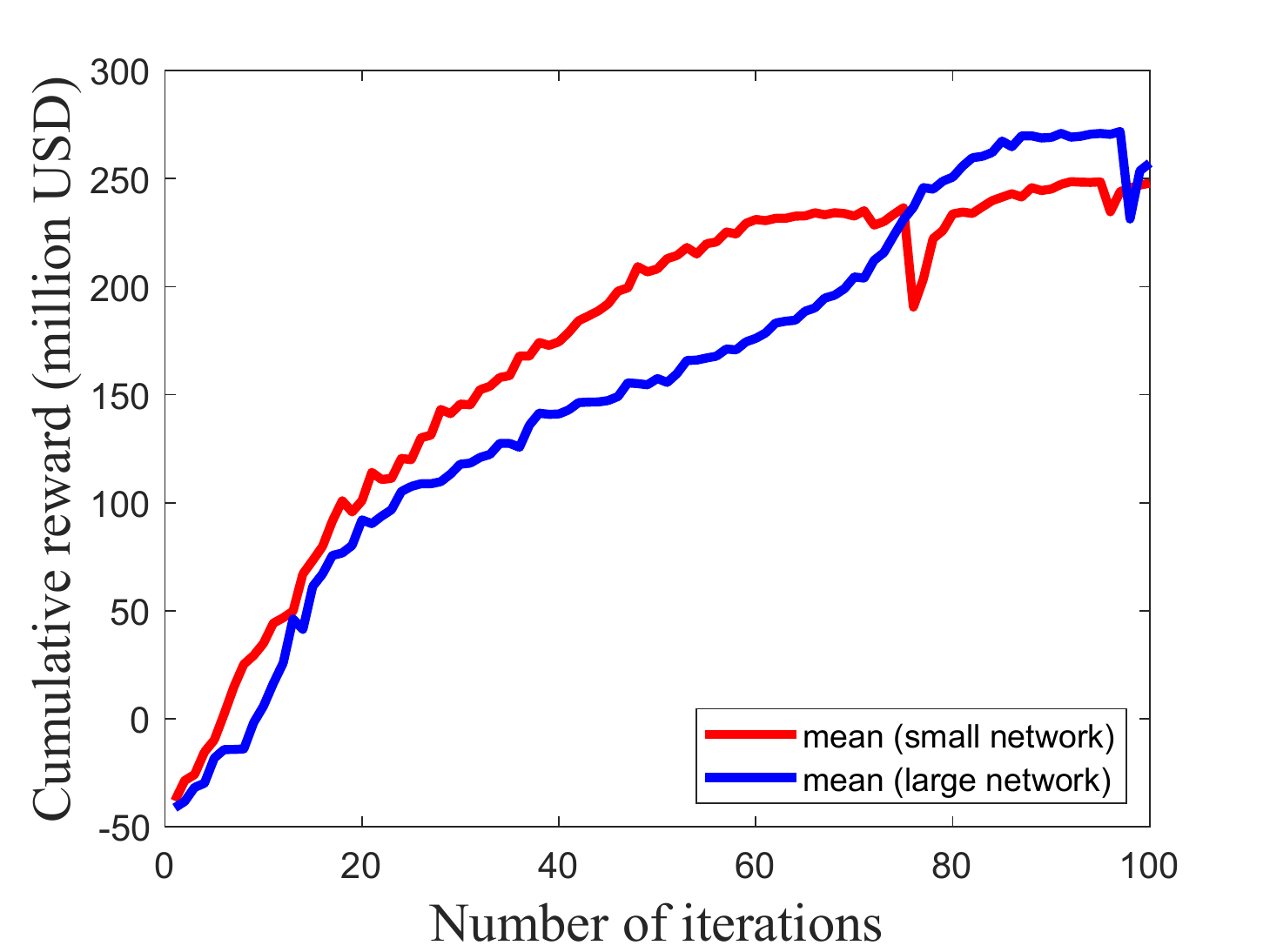}
\caption{Evolution of the mean cumulative reward with number of iterations}
\label{fig:reward}
\end{figure}

\section{Results}

The results for the optimization problem using the small and large CNN architectures are now presented. Fig. \ref{fig:reward} shows the evolution of the cumulative reward with number of iterations. The red and blue curves shows the evolution of the mean reward for the small and large networks, respectively. From the result it is evident that the agent is learning how to perform the task due to the consistent increase in the mean reward. Additionally, the mean reward for both networks at later stage of the optimization has about same value as the maximum reward. The result shows that the small network has a relatively faster convergence rate compared to the large network, but the large network obtained the highest cumulative reward.

\begin{figure}[!htb]
    \centering
    \subfigure[Small network ($\$249.80$ million)] {\label{fig:sol_small}\includegraphics[width=0.4\textwidth]{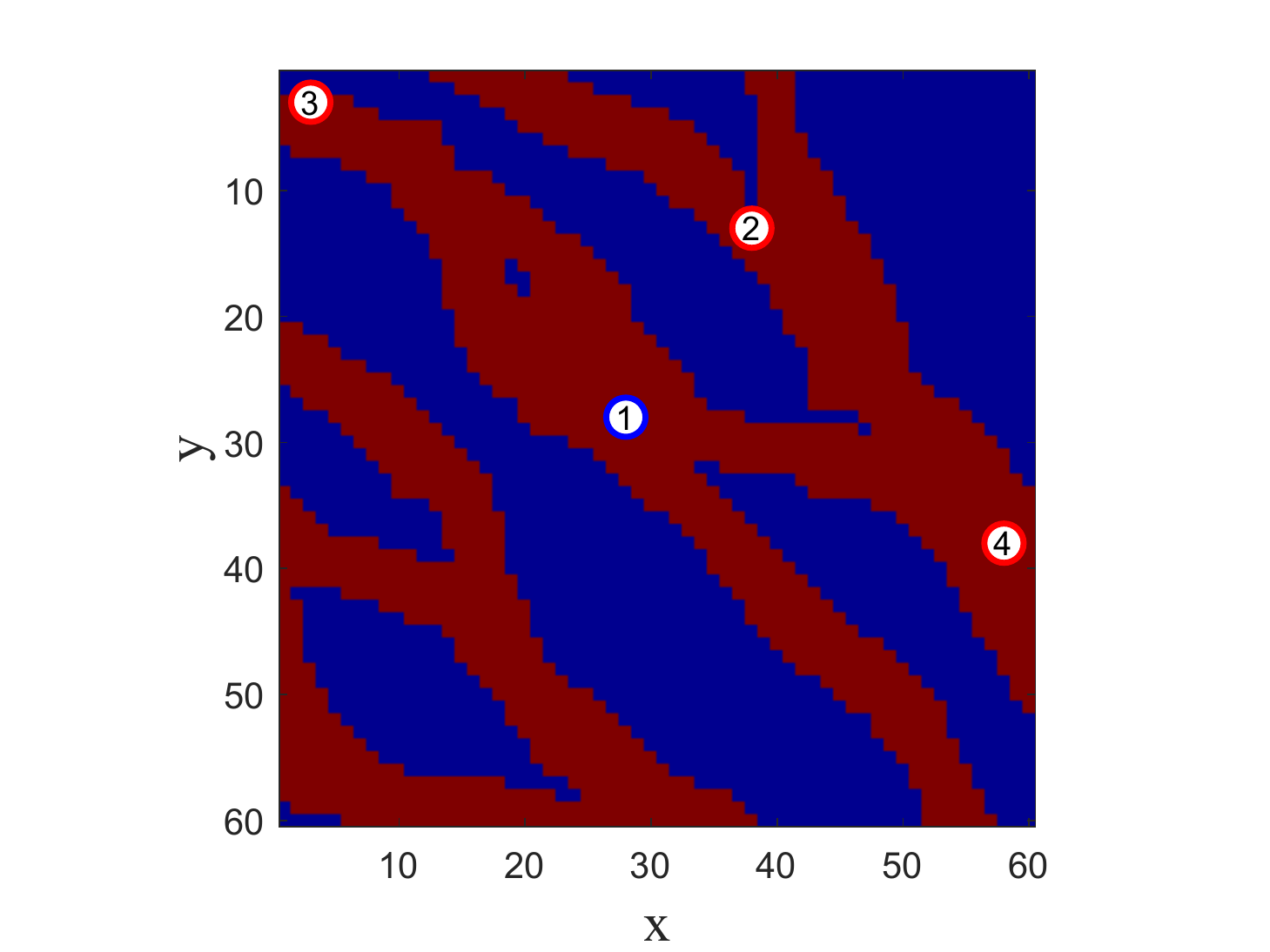}}
    \subfigure[Large network ($\$271.86$ million)] {\label{fig:sol_large}\includegraphics[width=0.4\textwidth]{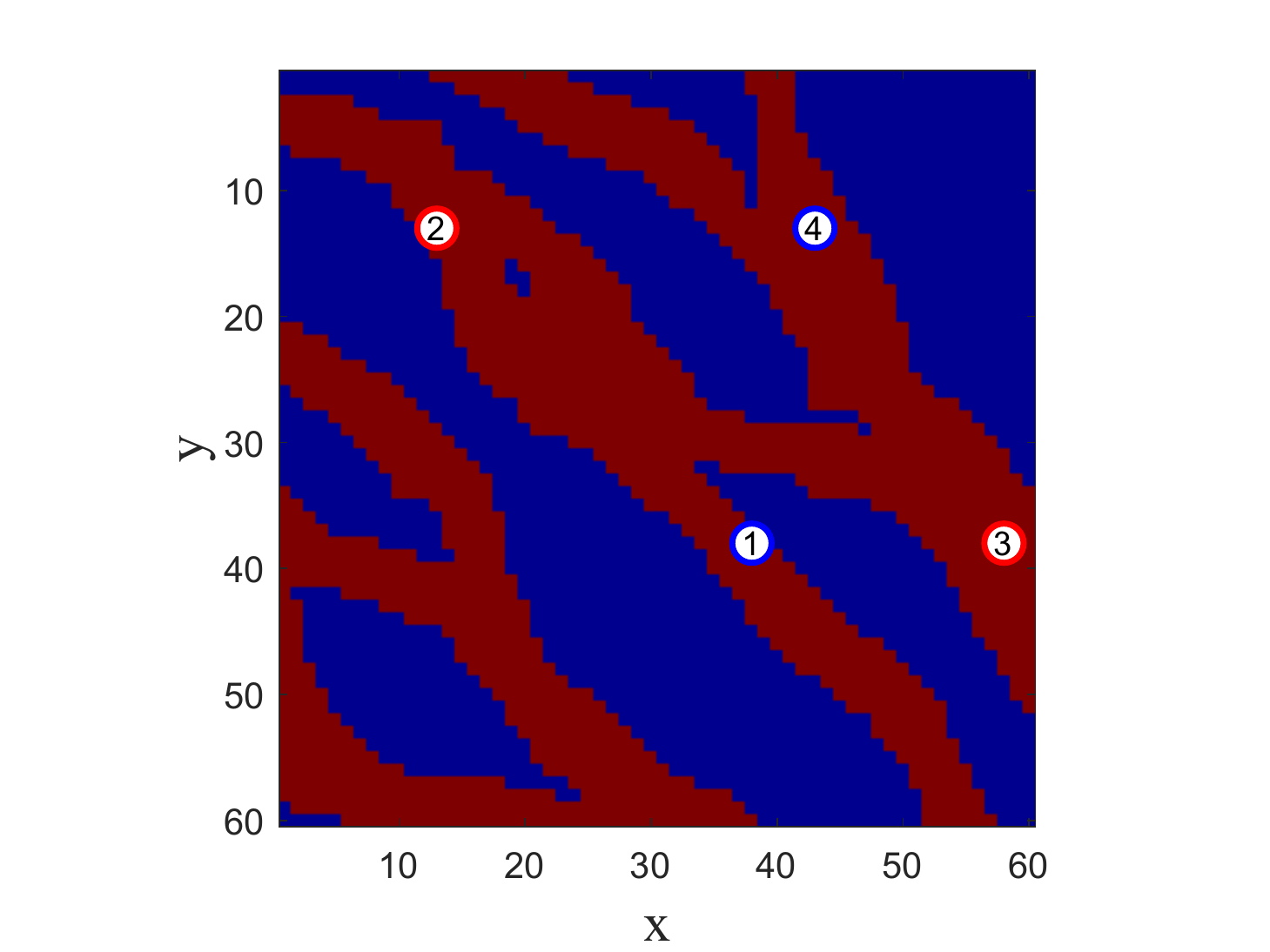}}
	\subfigure[PSO-MADS ($\$272.12$ million)] {\label{fig:sol_pso}\includegraphics[width=0.4\textwidth]{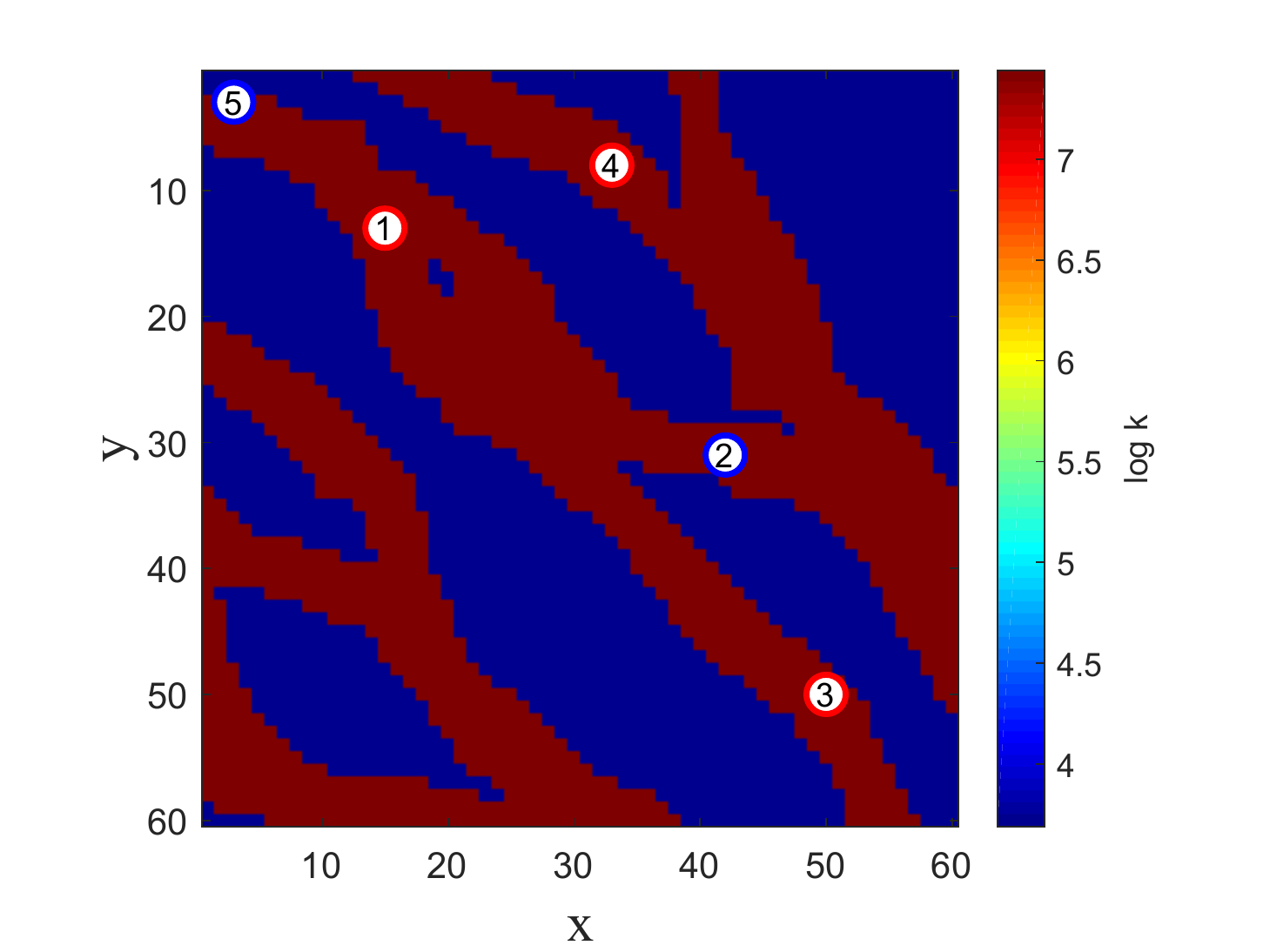}}
    \caption{Optimal solutions (with cumulative reward, NPV) obtained by following the policy of the small and large network. The best PSO-MADS solution from three runs is also shown. The blue circles represent injectors, red represent producers, and the number denotes the drilling stage in which the well is drilled.}
	\label{fig:opt_sol}
\end{figure}

PSO-MADS algorithm is also used to benchmark the performance of PPO. The optimization consists of 15 decision variables, a PSO population size of 40 and a termination criteria of 10,000 flow simulations. The optimization is fully  parallelized using 40 CPUs. As noted earlier, using PSO-MADS we obtain a single solution (and not a policy) that (potentially) maximizes the NPV. Fig.~\ref{fig:sol_small}~and~\ref{fig:sol_large}  shows the 'optimal' solutions obtained by following the policy of the small and large networks, respectively. The results for PSO-MADS is shown in Fig. \ref{fig:sol_pso}. Although, the DRL-based solutions obtained similar number of wells, the solution of the small network entails a development with three producers and one injector, while the large network's solution includes two injectors and two producers. In both solutions, however, an injector is drilled in the first stage.

PSO-MADS obtained a solution with a cumulative reward of \$272.12 million. The cumulative reward obtained by both the small and large network policies are \$249.80 and \$271.86 million, respectively. Although PSO-MADS obtained a higher cumulative reward than that obtained by following the policy of the large network, the PSO-MADS solution includes a total of five wells while that of the large networks solution has four wells (less capital cost) and comparable NPV. This results suggests that the policy obtained by DRL are applicable for field development optimization. 

\noindent
\\
\textbf{Policy generalization}
\\

It is evident from the results that PSO-MADS have a lower computational cost than the DRL approach. This is mainly because PSO-MADS solves a relatively simple problem of finding the optimal solution that maximizes the NPV, while in DRL our goal is to find a policy that maps from varying states to optimal actions. Thus, the main goal of applying DRL to field development optimization is to obtain a policy that is robust towards small changes in the problem setup. The generalization of the policy obtained from the large network is now investigated.

\begin{figure}
    \centering
    \includegraphics[width=0.4\textwidth]{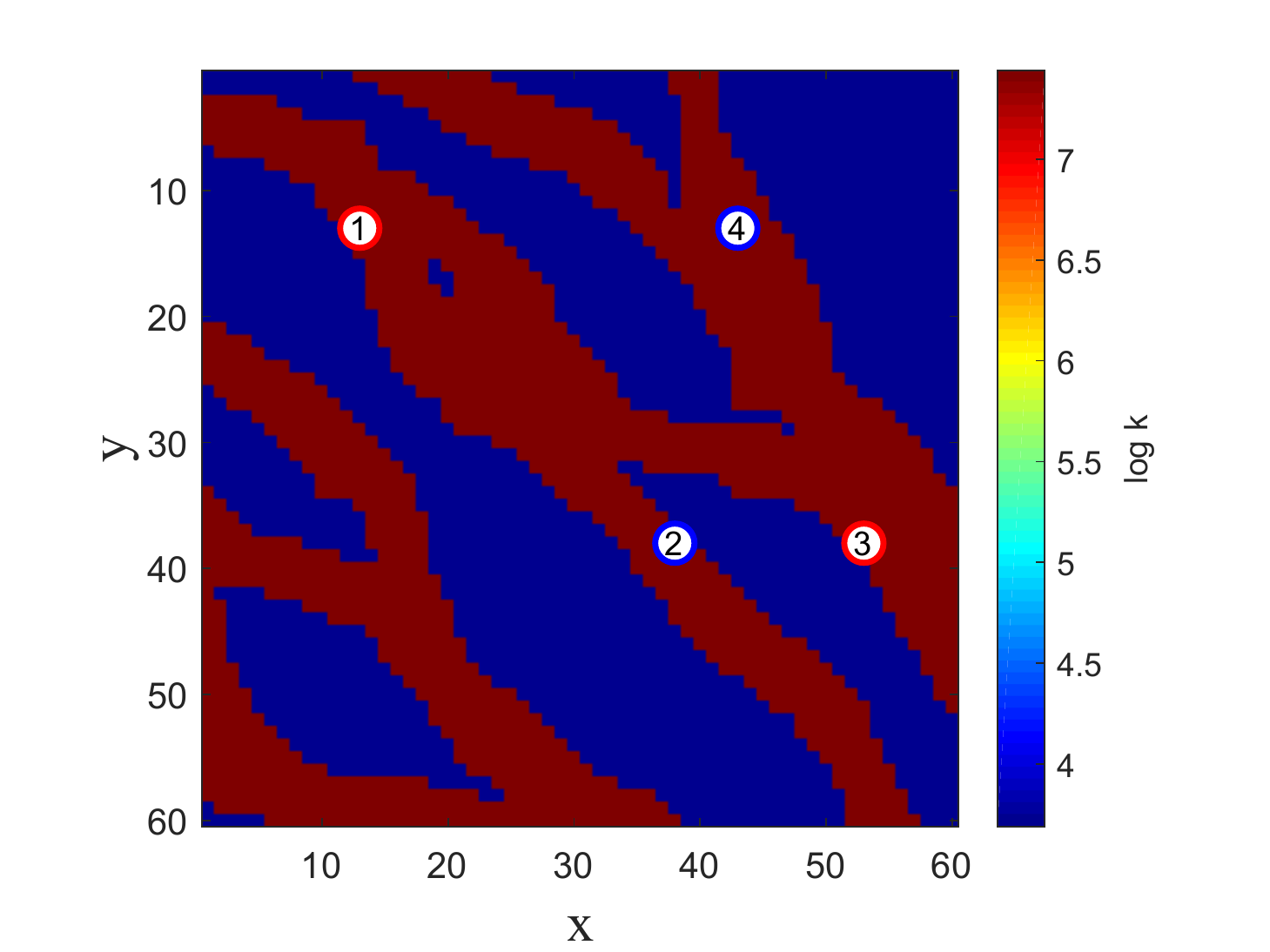}
    \caption{Optimal solutions (with an NPV of \$269.06 million) obtained by following the policy of the large network after user defined initial action.}
	\label{fig:sol_gen}
\end{figure}

The action proposed by the policy of the large network in the second drilling stage, which is to drill a producer (in the northwestern region) is executed in the first stage. This means the state after the first drilling stage will be different from what would have been obtained if an injector is drilled first as recommended by the policy. Interestingly, in the PSO-MADS solution the first action involves a producer drilled in approximately same area in the northwestern region. After choosing this first action, the policy of the large network is followed from the second to fifth stage. The result obtained is shown in Fig. \ref{fig:sol_gen} shows the solution obtained with an NPV of \$269.06 million. The policy maintained the location of the subsequent wells (except the slight alteration of the location of the producer drilled in the third drilling stage), while drilling an injector in drilling stage two as will be expected.

\section{Conclusion}

In this work, CNN-based DRL algorithm was applied to the field development optimization problem. Two different CNN architectures of varying layers and composition were considered. The proximal policy optimization (PPO) \cite{schulman2017proximal} deep reinforcement learning algorithm was considered. The performance of PPO was bench marked with a hybrid derivative-free PSO-MADS \cite{isebor2014a} optimization algorithm that has been shown to be effective for the field development optimization problem. 

The policy obtained by using the large network considered in this work provided better solution than that obtained from the policy of the small network. However, the small network has a faster convergence rate which may be due to the fewer number of parameters in the network. Preliminary investigation of the generalization ability of the policy obtained was successful.

In future work, the sampling efficiency of the DRL algorithm should be improved by tuning both the PPO and neural network hyperparameters. The performance of other DRL algorithms, such as soft actor critic (SAC) \cite{haarnoja2018soft}, importance weighted actor-learner architectures (IMPALA) \cite{espeholt2018impala} (and PPO variant of IMPALA), distributed prioritized experience replay (APEX) \cite{horgan2018distributed}, and model-based reinforcement learning that have demonstrated great sample and computational efficiency in other domains, should be investigated on the field development optimization problem. The generalization of the DRL policy should also be further investigated on cases with varying model parameters. It will be of interest to also incorporate varying well operational settings into the DRL policy by including an additional well settings decision variable during training. Finally, the DRL algorithm should be extended to field development under uncertainty where several realizations of the reservoir model are used to capture geological uncertainty.

\noindent
\\
\textbf{Acknowledgement}
\\

Thanks to Meng Tang for directing me to the RLlib package used in this work. The computational resources provided by Stanford's Center for Computational Earth and Environmental Sciences (CEES) is greatly appreciated.

\FloatBarrier

{\small
\bibliographystyle{ieee_fullname}
\bibliography{egbib}
}

\end{document}